\documentclass[aip,reprint]{revtex4-1} 

\usepackage{graphicx}
\usepackage{dcolumn}
\usepackage{bm}
\usepackage{natbib}
\usepackage{textcomp}
\usepackage{dsfont}
\usepackage{amsmath}
\usepackage{amssymb}
\usepackage{color}
\usepackage{bbm}

\newcommand{\IsoSi}{$^{28}$Si}
\newcommand{\Phos}{$^{31}$P}

\begin{document}

\title{{NMR study of optically hyperpolarized phosphorus donor nuclei in silicon}}

\author{P. Gumann} \email{gumann@us.ibm.com} 
\altaffiliation{These authors contributed equally to this work.}
\affiliation{IBM T.J. Watson Research Center, Yorktown Heights, NY 10598,  USA}
\affiliation{Institute for Quantum Computing, University of Waterloo, Waterloo, Ontario N2L 3G1, Canada}
\affiliation{Department of Physics and Astronomy, University of Waterloo, Waterloo, Ontario N2L 3G1, Canada}
\author{H. Haas} \email{hhaas@uwaterloo.ca}
\altaffiliation{These authors contributed equally to this work.}
\affiliation{Institute for Quantum Computing, University of Waterloo, Waterloo, Ontario N2L 3G1, Canada}
\affiliation{Department of Physics and Astronomy, University of Waterloo, Waterloo, Ontario N2L 3G1, Canada}
\author{S. Sheldon}
\affiliation{IBM TJ Watson Research Center, Yorktown Heights, NY 10598, USA}
\affiliation{Institute for Quantum Computing, University of Waterloo, Waterloo, Ontario N2L 3G1, Canada}
\author{L. Zhu}
\affiliation{Department of Physics and Astronomy, Dartmouth College, Hanover, NH 03755, USA}
\author{M.L.W. Thewalt} 
\affiliation{Department of Physics, Simon Fraser University, Burnaby, British Columbia V5A 1S6, Canada}
\author{D.G. Cory}
\affiliation{Institute for Quantum Computing, University of Waterloo, Waterloo, Ontario N2L 3G1, Canada}
\affiliation{Department of Chemistry, University of Waterloo, Waterloo, Ontario N2L 3G1, Canada}
\affiliation{Perimeter Institute for Theoretical Physics, Waterloo, Ontario N2L 2Y5, Canada}
\author{C. Ramanathan} \email{chandrasekhar.ramanathan@dartmouth.edu}
\affiliation{Department of Physics and Astronomy, Dartmouth College, Hanover, NH 03755, USA}

\date{\today}

\begin{abstract}
We use above-bandgap optical excitation, via a 1047~nm laser, to hyperpolarize the $^{31}$P spins in low-doped (N$_D =6\times10^{15}$~cm$^{-3}$) natural abundance silicon at 4.2~K and 6.7~T, and inductively detect the resulting NMR signal.  The $30$~kHz spectral linewidth observed is dramatically larger than the 600~Hz linewidth observed from a $^{28}$Si-enriched silicon crystal.  We show that the observed broadening is {consistent} with previous ENDOR results showing discrete isotope mass effect contributions to the donor hyperfine coupling.   A secondary source of broadening is likely due to variations in the local strain, induced by the random distribution of different isotopes in natural silicon.  {The nuclear spin T$_1$ and the build-up time for the optically-induced $^{31}$P hyperpolarization in the natural abundance silicon sample were observed to be $178\pm47$~s and $69\pm6$~s respectively, significantly shorter than the values previously measured in \IsoSi-enriched samples under the same conditions.   We also measured the T$_1$ and hyperpolarization build-up time for the $^{31}$P signal in natural abundance silicon at 9.4~T to be  $54\pm31$~s and $13\pm2$~s respectively.  The shorter build-up and nuclear spin T$_1$ times at high field are likely due to the shorter electron-spin T$_1$, which drives nuclear spin relaxation via non-secular hyperfine interactions}.  At 6.7~T, the phosphorus nuclear spin T$_{2}$ was measured to be $16.7\pm1.6$~ms at 4.2~K, a factor of 4 shorter than in \IsoSi-enriched crystals.  This was observed to further shorten to $1.9\pm0.4$~ms in the presence of {the} infra-red laser.

\end{abstract}
\maketitle

\noindent 
Phosphorus-doped silicon (Si:P) is a technologically important material in quantum applications \cite{Kane-1998,Ladd-2002,Stegner-2006,Morello-2015,Lo-2015}, as the donor spins have some of the longest coherence times observed for any solid-state spin system \cite{Tyryshkin-2006,Tyryshkin-2011}.  The growth of isotopically-enriched \IsoSi~crytals, where local magnetic field fluctuations due to the $^{29}$Si are eliminated, has enabled dramatically longer electronic\cite{Tyryshkin-2011} and nuclear\cite{Saeedi-2013} donor spin coherence times.  

Natural silicon consists of 3 isotopes: $^{28}$Si, $^{29}$Si and $^{30}$Si, whose relative abundances are 92.23\%, 4.67\% and 3.1\%\, respectively.  While $^{29}$Si is a  spin-1/2 nucleus, $^{28}$Si and $^{30}$Si are spin-0 nuclei.   In addition to suppressing spin-induced magnetic field noise, isotope engineering of silicon --- originally enabled by the Avogadro Project \cite{Andreas-2011} --- has improved our understanding of silicon physics.  Photoluminescence \cite{Karaiskaj-2002,Karaiskaj-2003} and ESR \cite{Tezuka-2010,Stegner-2010,Stegner-2011} experiments on boron-doped \IsoSi~and natural silicon have shown that the random distribution of silicon isotopes causes local changes to the valence band in the vicinity of the boron acceptor. The broad EPR lines observed at low doping concentrations in natural silicon were attributed to a distribution of local strain fields induced by the random spatial distribution of different isotopes \cite{Stegner-2011}. For shallow group-V donor states, changes in the electron-nuclear hyperfine interaction have been observed with electron-nuclear double resonance (ENDOR) to correlate with {the} host {crystal} isotope mass {distribution} \cite{Sekiguchi-2014}, though the {microscopic mechanisms underlying this effect are still unclear}.

Directly studying the spin properties of phosphorus nuclei at low donor concentrations has been a challenge. Given the low sensitivity of NMR measurements, direct inductive detection of phosphorus nuclear spins in silicon has previously only been possible at very high doping concentrations ($\sim10^{18}$~cm$^{-3}$) \cite{Alloul-1987,Jeong-2009}. {We recently demonstrated a direct inductive readout of the phosphorus NMR signal from an isotopically-enriched, \IsoSi-sample with a \Phos~donor concentration in the range of $10^{15}$~cm$^{-3}$, following hyperpolarization with a non-resonant infra-red laser \cite{Gumann-2014}}.  The inductively detected NMR data complemented results from optically\cite{Steger-2011} and electrically\cite{Saeedi-2013} detected experiments on \Phos~at similar dopant concentrations.  {Here, we utilize optical hyperpolarization and direct inductive readout to characterize the properties of \Phos~spins in a natural silicon crystal, and contrast the results with data from \IsoSi-enriched samples.}

\begin{figure}[!bth]
\includegraphics[scale=0.4]{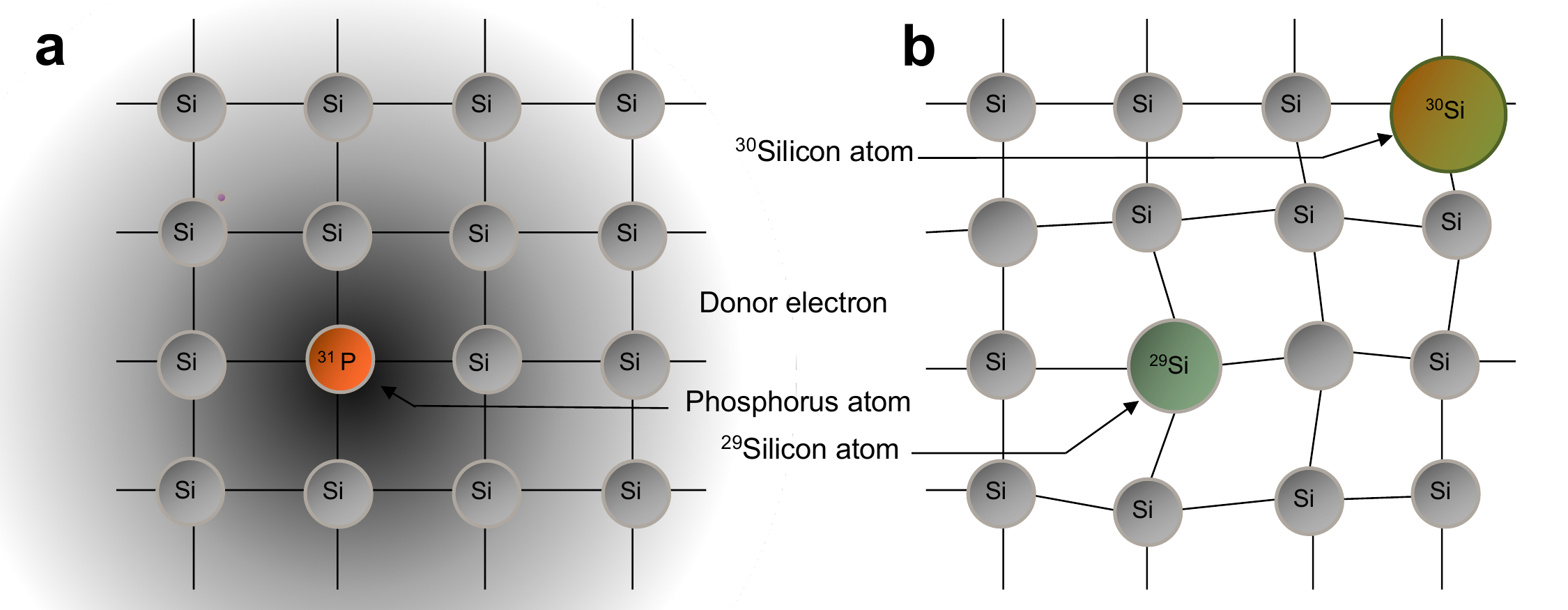}
\centering
\includegraphics[scale=0.35]{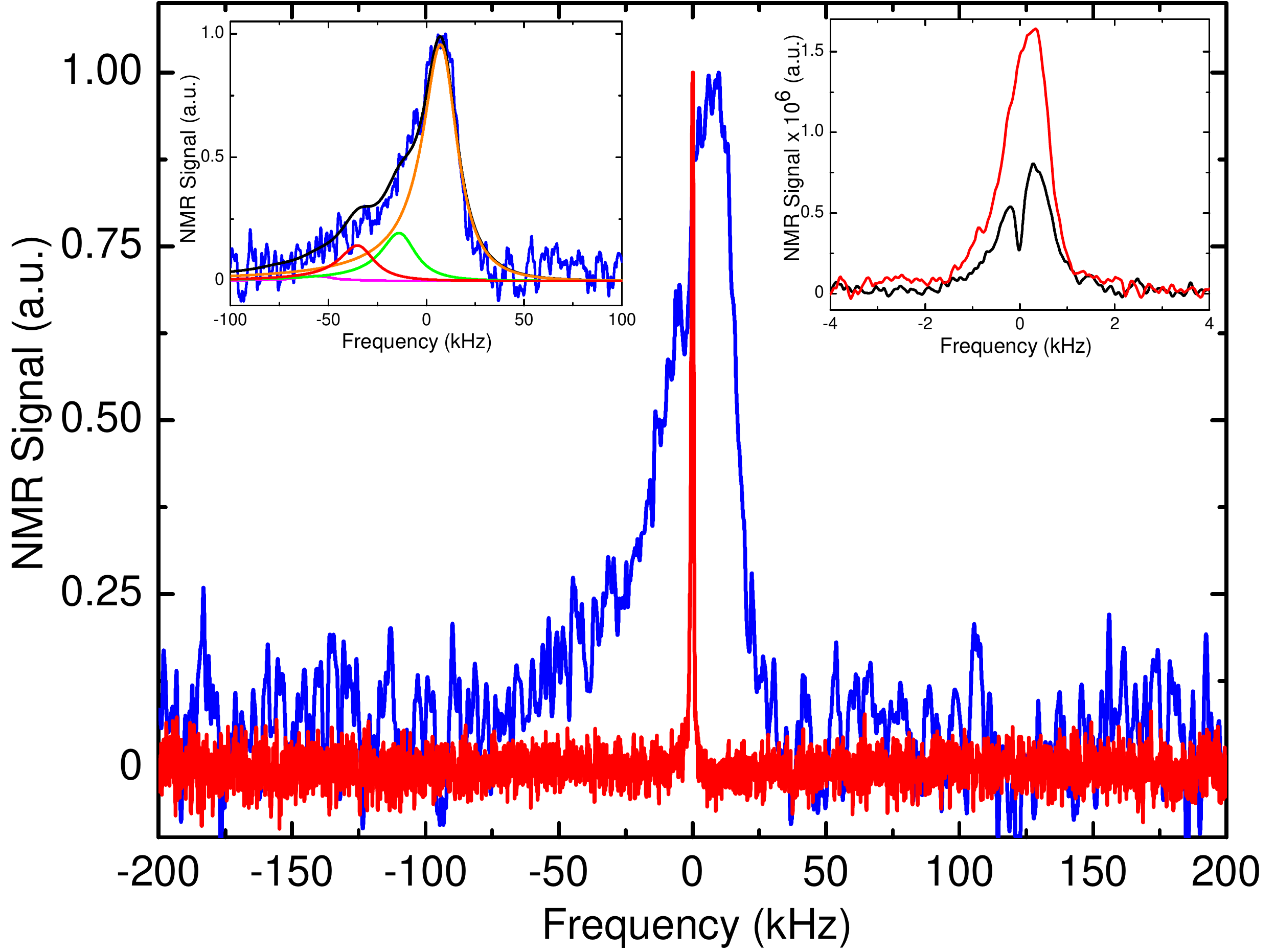}
\caption{Schematic representation of (a) a uniform $^{28}$Si-enriched crystal, and (b) a natural silicon lattice containing $^{28}$Si, $^{29}$Si and $^{30}$Si atoms. The 2~nm Bohr radius of the shallow \Phos~donor electron extends over a large number of silicon lattice sites. The main plot below shows normalized \Phos~NMR spectra from \IsoSi~(red) and natural silicon (blue) samples, measured at 6.7~T and 4.2~K. {A 1 kHz exponential line broadening has been applied to the spectrum from the natural silicon sample and a 100 Hz line broadening to the spectrum from the \IsoSi~sample.} The left inset shows a comparison of the observed NMR spectrum for natural silicon (shown in blue) and a simulation of the lineshape (shown in black) expected from the mass effect model \cite{Sekiguchi-2014}. The inset also shows the 4 largest relative contributions from {$M_{\text{NN}}\approx28$} (orange), {$M_{\text{NN}}\approx28.25$} (green), {$M_{\text{NN}}\approx28.5$} (pink), and {$M_{\text{NN}}\approx28.75$} (red).
The right inset shows the result of an NMR hole-burning experiment on the \IsoSi~sample, with a 10~s long saturation pulse using a weak 3~Hz Rabi frequency, indicting the inhomogeneous nature of the spectral line. The width of the spectral hole is about 200~Hz. 
}
\label{fig:fig1}
\end{figure}

Figure~\ref{fig:fig1} shows the normalized NMR spectra obtained from \IsoSi~(red) and natural silicon (blue) samples. The {spectrum from the \IsoSi~sample} was recorded with 16 averages, and an optical hyperpolarization time of 400~s, whereas the {spectrum from the natural silicon sample} was acquired with 256 averages, with an optical hyperpolarization time of 250~s.  
Both spectra were obtained using a commercial Bruker Avance NMR Spectrometer at 6.7~T and $4.2\pm0.3$~K.   Under these conditions, the thermal electron spin polarization is 79~\%. We probed the nuclear spins in the lower electron spin manifold, with a transition at 174~MHz, where the thermal nuclear spin polarization is 0.1~\%. The length of the $\pi/2$ pulse {used} was 2.5~$\mu$s. {The optical excitation was performed with a 150~mW, 1047~nm, above-bandgap laser, with a linearly polarized beam of {9~mm} effective size. The (indirect) bandgap in silicon is 1.17~eV which corresponds to an optical wavelength of 1059~nm \cite{Bludau-1973}. The penetration depth for 1047~nm light in silicon at cryogenic temperatures is a few centimetres which allowed the excitation of bulk phosphorus {donors throughout the sample} \cite{Macfarlane-1959}.} 

Similar sized \IsoSi~and natural silicon samples, measuring $2\times2\times8~\text{mm}^3$, were mounted in a strain-free configuration. {The \IsoSi~sample was a dislocation-free crystal \cite{Becker-2010} with a phosphorus donor concentration of $1.5 \times 10^{15}$ cm$^{-3}$, while the natural silicon sample was a float-zone grown commercial silicon sample (Topsil) with a phosphorus donor concentration of $6 \times 10^{15}$ cm$^{-3}$.  The boron concentration in both samples was less than $1.0 \times 10^{14}$ cm$^{-3}$.}
{Both samples were etched in HF/HNO$_3$ before the experiments. We also confirmed the same natural silicon linewidth after annealing and etching another separate sample cut from the same crystal.} In each case, the samples were placed in a silver plated copper, \textit{RF}-coil, connected to a resonant, low temperature \textit{LC}-circuit. Home-built NMR probes were then immersed into commercial, liquid helium dewars with a set of optical, sapphire windows located at the bottom of each dewar, and aligned with the main B$_{0}$ magnetic field of both magnets. 

The \Phos-spectra from the \IsoSi-enriched sample is observed to have a linewidth of $600$~Hz, while the linewidth of the natural silicon crystal is substantially broader, about $30$~kHz. Such considerable changes have previously been observed in ESR-detected ENDOR spectra of Si:P samples with varying silicon isotopic concentration. In order to understand the difference in the observed linewidths, it is useful to consider the effective Hamiltonian for an isolated phosphorus donor impurity at high magnetic field:
$$
\mathcal{H} = - \gamma_n B_z I_z - \gamma_e B_z S_z + \frac{2\pi}{\hbar} A S_z I_z + \mathcal{H}_{n}^{d}
\label{eq-Ham} ,
$$
where $\gamma_n/2\pi = 17.23$~MHz/T and $\gamma_e/2\pi = -28.024$~GHz/T are the nuclear and electron gyromagnetic ratios, $A=117.5~\text{MHz}$ is the nominal strength of the isotropic hyperfine interaction, and $\mathcal{H}_{n}^{d}$ represents the magnetic dipolar coupling between the phosphorus nucleus and other electronic and nuclear spins in the system. 
At high field the eigenstates are almost exactly given by the product states $\mid \uparrow_{e} \uparrow_{n}\rangle$, $\mid \uparrow_{e} \downarrow_{n}\rangle$, $\mid \downarrow_{e} \uparrow_{n}\rangle$, $\mid \downarrow_{e} \downarrow_{n}\rangle$ \cite{Schweiger-2001}.  The resonance frequency for a nuclear spin transition is given by $\gamma_n B_z \pm \pi A$, corresponding to frequencies of $174.08$~MHz and $56.58$~MHz respectively. As noted above, the experiments shown here were performed on the larger $174.08$ MHz transition. 
At low doping concentrations, the largest contribution to $\mathcal{H}_{n}^{d}$ in natural silicon is the phosphorus-silicon dipolar coupling, which we estimate to be on the order of 200 Hz in a natural abundance crystal -- about twice the strength of the observed silicon-silicon dipolar linewidth in natural silicon\cite{Christensen-2017}. This is insufficient to explain the observed broadening.

It is apparent that any variation in the hyperfine strength $A$ will cause a shift in the observed NMR line.  The hyperfine interaction strength $A$ is given by 
$$
A = \frac{2}{3}\frac{\mu_0}{\hbar}\gamma_e\gamma_n |\psi(0)|^2
$$
where $\psi(0)$ is the magnitude of the electronic wavefunction at the \Phos~nucleus \cite{Schweiger-2001}.  Isotope variations can result in both local symmetry breaking \cite{Wilson1961} as well as change the effective mass of the electron \cite{Sekiguchi-2014}, resulting in changes to $\psi(0)$ and the hyperfine interaction strength.  In the previous ENDOR experiments, the shape of the observed spectra were seen to strongly correlate with the isotopic composition of the host lattice, and the authors extracted the following relation for the variation of hyperfine interaction strength(s) with isotopic composition: \cite{Sekiguchi-2014}  
$$
A = A_{28} + \alpha_{\text{NN}}\left(M_{\text{NN}}-M_{28}\right) + \alpha_{\text{bulk}}\left(M_{\text{bulk}}-M_{28}\right),
$$
where $A_{28}$ is the \Phos~hyperfine interaction in a \IsoSi-enriched lattice, $M_{\text{NN}}$ is the average mass of the four nearest-neighbor silicon isotopes, $M_{28}$ is the mass of \IsoSi~and $M_{\text{bulk}}$ is the average bulk {isotopic} mass. 

For the \IsoSi~samples, there should be little to no variation in $A$, as observed in the experiment.  The nuclear spin T$_2$ for the \Phos~in \IsoSi~has been measured to be 56 ms {\cite{Gumann-2014}}, suggesting that most of the 600 Hz broadening arises from local magnetic field inhomogeneities.  The right inset shows the result of an NMR hole-burning experiment with a 10 s long saturation pulse using a weak 3 Hz Rabi frequency.  The width of the spectral hole is about 200 Hz.  {It should be noted} {that} {the relative intensities for the two spectra in this inset are arbitrary as the experimental parameters, such as polarization time and sampling periods were optimized independently for each spectra.}

The left inset to Figure~1 shows a comparison of the observed NMR spectrum for natural silicon (shown in blue) and a simulation of the lineshape (shown in black) expected from this model. {Zero and first order phase corrections to the experimental data were adjusted to yield the best fit to the model lineshapes.  The natural silicon spectum was also shifted by $-1$~kHz to optimize the fit.}  {This} {shift is consistent with small} {frequency} {shifts that we observe for the \IsoSi-enriched sample spectra between different experimental runs, which happen due to} {changes in sample positioning.}  The model uses the experimentally measured \cite{Sekiguchi-2014} parameters $\alpha_{\text{NN}} = -170$ kHz/u and $\alpha_{\text{bulk}} = 117$ kHz/u, and the isotopic masses in natural silicon to find the peak centers. {Each peak is convolved with the same anisotrospic Lorentzian lineshape function \cite{Stancik-2008} used by Sekiguchi \emph{et~al.} \cite{Sekiguchi-2014}, with a linewidth of 22.45~kHz and asymmetry parameter $3.2 \times 10^{-5}$}. 

There are 9 possible values for $M_{\text{NN}}$ starting from {$M_{\text{NN}} = M_{28} = 27.9769265325$ u \cite{Audi-2003}, and increasing in steps of 0.25 u to $M_{\text{NN}}\approx 30$ u}.  
The inset also shows the 4 largest relative contributions from {$M_{\text{NN}}\approx28$} (orange), {$M_{\text{NN}}\approx28.25$} (green), {$M_{\text{NN}}\approx28.5$} (pink), and {$M_{\text{NN}}\approx28.75$} (red). There is good agreement for both the center of the line {with respect to the \IsoSi~data}, and the width of the spectrum. 

The residual broadening captured by {the anisotrospic Lorentzian lines is} most likely due to strain-induced hyperfine changes due to the random isotope distribution, which were originally studied by Wilson and Feher \cite{Wilson1961}, and recently re-explored by Mansir {\em et al}.\cite{Mansir-2018} The random distribution of different isotopes has been observed to result in a broad distribution of strain fields in boron-doped silicon ESR studies \cite{Stegner-2011}.  

We also compared the properties of the optically-induced \Phos~NMR signal at two different magnetic fields, 6.7~T and 9.4~T.  At 9.4~T and 4.2~K the equilibrium electron spin polarizations is 91\%.  The NMR experiments were once again performed on the larger 220.7 MHz transition which has a thermal equilibrium polarization of 0.13\%.  
The experiments at 9.4~T used 3 different pieces from the natural silicon crystal, while the build-up and relaxation experiments at 6.7~T used two pieces from the same crystal.  A similar broad NMR line was observed at 9.4~T (data not shown).

\begin{figure}[!bth]
\centering
\includegraphics[scale=0.4]{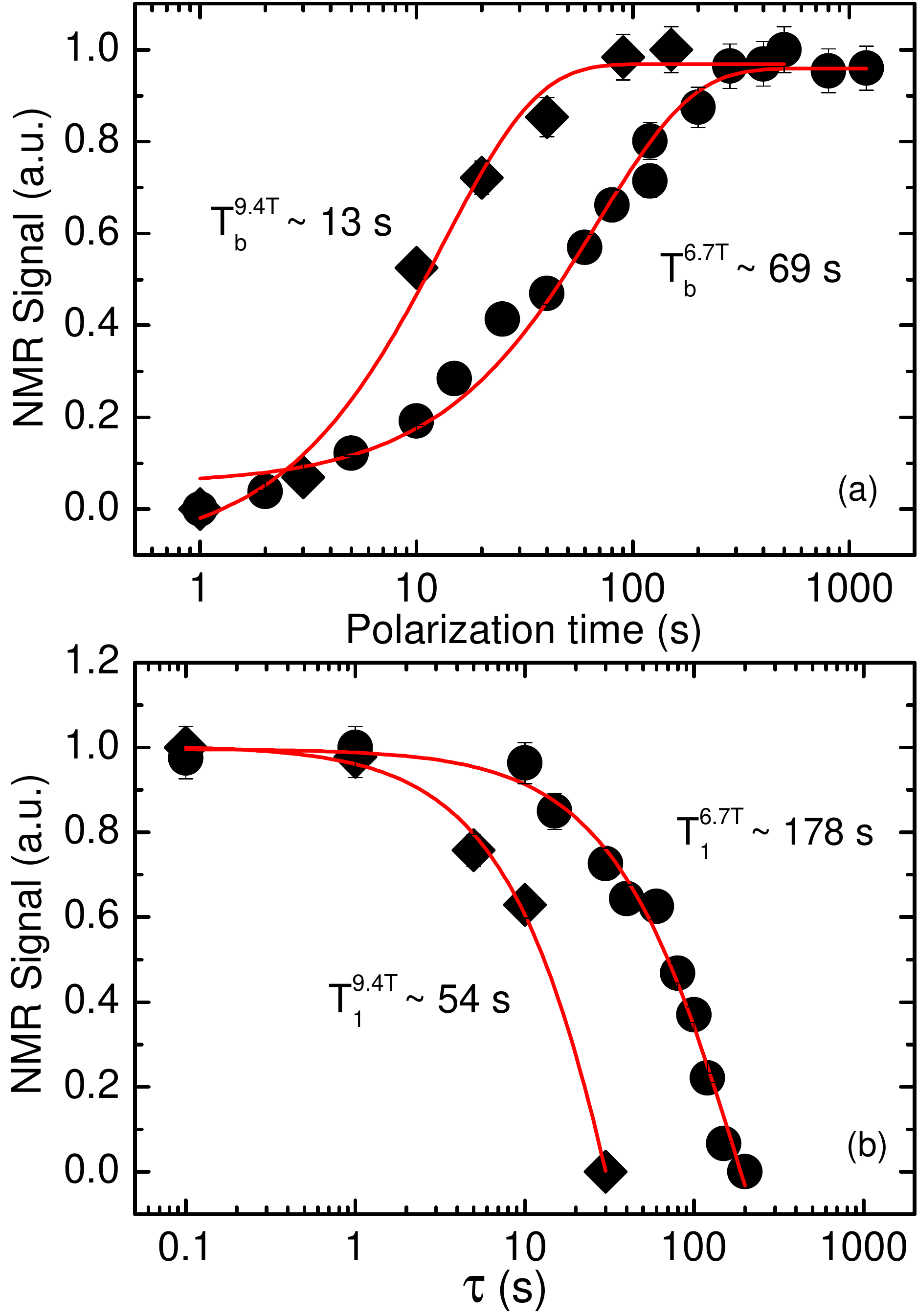}
\caption{
(a) Build-up of the hyperpolarized $^{31}$P signal at 9.4~T (black diamonds) and 6.7~T (black circles) in natural silicon as a function of illumination time ($\tau$) using a 1047 nm laser. The normalized data were fit to the function: $1- \exp\left(\tau/T_b\right)$, to find the characteristic build-up times T$_b^{9.4T}$=13\textpm2~s, and T$_b^{6.7T}$=69\textpm6~s. (b) Decay of the optically hyperpolarized $^{31}$P signal in natural silicon at 9.4~T (black diamonds) and 6.7~T (black circles), yielding T$_1$ relaxation times of T$_1^{9.4T}$=54~s and T$_1^{6.7T}$=178~s respectively. {For both subplots, the maximum signal at the two magnetic fields has been normalized to have a value of one since our experiments did not provide the magnitude of spin polarization.}
}
\label{fig:fig2}
\end{figure}

{Figure~\ref{fig:fig2}(a) shows the build-up of the $^{31}$P-spin polarization accomplished by illuminating the sample with a lower power 100~mW, 1047~nm laser.} The build-up curves of the optical hyperpolarization were measured by increasing the laser excitation time (or polarization time), from 1~s to 16 minutes for the 6.7~T data set, and from 1~s to 100~s for the 9.4~T results. We were able to fit the build-up curves using a single exponential with characteristic times of $13\pm2$~s at 9.4~T and $69\pm6$~s at 6.7~T fields. {The thermal equilibrium polarization could not be measured, making it difficult to directly quantify either the sign or the magnitude of the nuclear spin polarization.}  {However, previous ESR measurements suggest a negative nuclear spin polarization is obtained following optical hyperpolarization \cite{McCamey-2009,Gumann-2014}. }

Figure~\ref{fig:fig2}(b) shows the results of T$_{1}$ relaxation measurements at the two fields. The data were recorded using a 200~s laser polarization pulse, followed by a delay time $\tau$, during which the laser was turned off, and a $\pi/2$-read-out pulse. Both data sets are fit with a single exponential decay, yielding T$_1^{6.7T}=178\pm47$~s and T$_1^{9.4T}=54\pm31$~s. While build-up and relaxation times typically scale together, build-up times for hyperpolarization schemes are generally shorter than relaxation times since build-up times combine the rate of driven polarization transfer with the rate of relaxation. The observed build-up times also depend on the optical coupling of the light to the sample in a given experiment, which could vary depending on laser alignment and sample positioning.

The observed build up and relaxation times  for the natural silicon samples are significantly shorter than the 577~s build-up time and 712 s {\Phos~nuclear} $T_1$ relaxation time measured at 6.7~T and 4.2~K for the \IsoSi-enriched sample \cite{Gumann-2014}. While the phosphorus concentration in the natural silicon samples is about 4 times higher, it is known that for donor concentrations below  $10^{16}$~cm$^{-3}$, the electron spin T$_1$ is independent of donor concentration, \cite{Tyryshkin-2003} at least at low magnetic fields.  At low  fields \IsoSi~isotopic enrichment does not appear to change the electron spin T$_1$ times either \cite{Tyryshkin-2003}.
However, it has also been demonstrated\cite{Dluhy-2015} that matching the hyperfine shifted phosphorus Larmor frequency to that of the $^{29}$Si Larmor frequency induces an efficient resonant spin polarization transfer from \Phos~to $^{29}$Si nuclei. In our experiment the $56.58$~MHz hyperfine shifted \Phos~resonance {is} only $50$~kHz shifted from the $^{29}$Si Larmor frequency and could therefore be yielding a shortened {T$_1$}.  Isotopic variations could also modify the phonon spectrum, which could be important in the optical hyperpolarization process.

{
The shorter build-up and nuclear T$_1$ times at high field are likely due to the shorter electron-spin T$_1$ at high field, which in turn, drives nuclear spin relaxation via non-secular hyperfine interactions \cite{Pines-1957}.  The direct electron spin-phonon relaxation process is expected to scale with temperature and field {as} $T_1^{-1} \propto B^5 \coth\left(\frac{\hbar\gamma B}{k_B T}\right)$. At 4.2~K, we expect $\hbar\gamma B \ll k_B T$, and $T_1^{-1} \propto B^4 T$.\cite{Honig-1958,Honig-1960} 
In the presence of light, the electron T$_1$ is known to be further shortened by up to two orders of magnitude driven by the creation of non-thermal resonant phonons, photoionization and photoneutralization due to exciton capture processes\cite{Haynes-1960,Thewalt-1977}, exchange interaction with photocarriers\cite{FeherGere-1959} and trapping and re-emission of electrons\cite{Morley-2008}, with T$_1$ {measured to be} on the order of 2~ms in the presence of light and almost 20~ms in the dark at 8.56~T \cite{Morley-2008,McCamey2-2012}. 
}

\begin{figure}[!bth]
\centering
\includegraphics[scale=0.35]{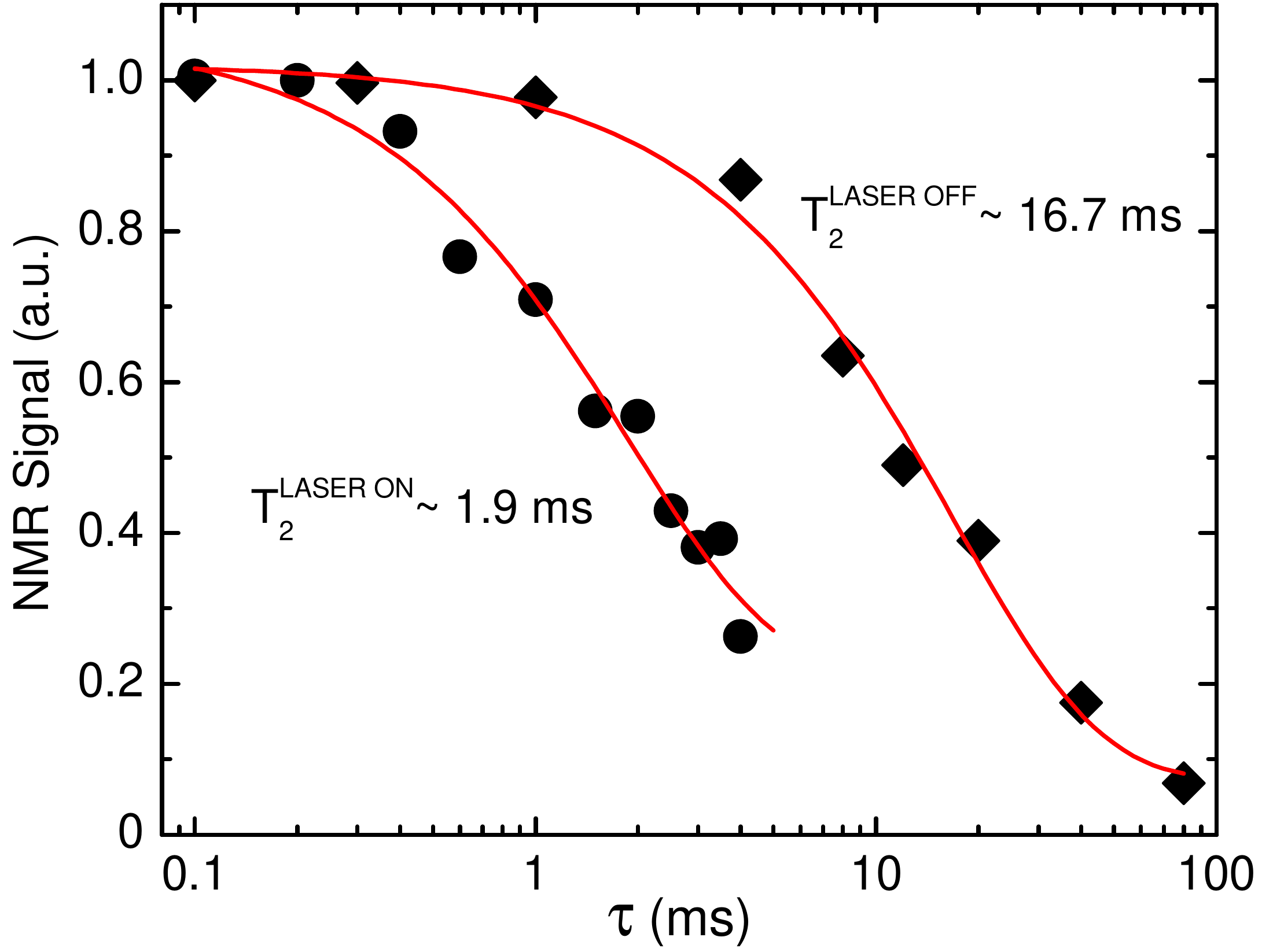}
\caption{\Phos~nuclear spin coherence time, T$_{2}$, in natural silicon measured with the Hahn echo at 4.2~K temperature, 6.7~T field. For the black circles the laser was kept ON during acquisition, T$_{2}=1.9\pm 0.4$~ms while for the red diamonds the laser was turned OFF during the acquisition, T$_{2}=16.7\pm 1.6$~ms. All data were measured with Hahn echo, with 150~s of optical polarization provided by a 1047~nm, and 150~mW, above-gap laser.
}
\label{fig:fig3}
\end{figure}

Figure~\ref{fig:fig3} shows the results of spin-echo experiments performed to measure the coherence time of the $^{31}$P nuclear spins, following 200~s of laser irradiation.  {The higher-power 150 mW 1047 nm laser was used for the optical excitation here.}  We measured the signal decay with the laser off and on, and fit the data to a single exponential decay to obtain nuclear spin T$_{2}$ values of 16.7~ms and 1.9~ms for the two cases respectively. 
In the absence of light the Hahn-echo T$_2$ in natural silicon is about a factor of 4 shorter than {the 56 ms T$_2$} observed with \IsoSi, due to the presence of the magnetically-active $^{29}$Si spins.  
The order of magnitude change in T$_2$ (1.9~ms to 16.7~ms) measured in the presence of light is likely due to the rapid modulation of the electron spins when the light is on, which results in local field fluctuations.

Non-resonant optical hyperpolarization of donor nuclei provides an important new tool to probe their local magnetic environment.  The dramatic difference in \Phos~NMR spectral linewidths measured for natural and \IsoSi~enriched silicon samples is suggested to arise from a combination of previously proposed isotope mass effects and strain-induced broadening due to the random distribution of isotopes in natural silicon.  At 6.7~T the build-up time for the optical hyperpolarization and the nuclear-spin T$_1$ for natural silicon are observed to be dramatically shorter than for \IsoSi.  In our experiments the $56.58$~MHz hyperfine shifted \Phos~resonance lies only $50$~kHz away from the $^{29}$Si Larmor frequency and could therefore be yielding a shortened $T_1$.  Isotopic variations could also modify the phonon spectrum, which could be important in the optical hyperpolarization process.
 The shorter build-up and nuclear T$_1$ times measured at 9.4~T are likely due to the shorter electron-spin T$_1$, which drives nuclear spin relaxation via non-secular hyperfine interactions.   At 6.7~T, the $^{31}$P-T$_{2}$ was measured to be $16.7\pm1.6$~ms at 4.2~K, a factor of 4 shorter than in \IsoSi~enriched crystals.

{This work was undertaken thanks in part to funding from the the Canada First Research Excellence Fund (CFREF). Further support was provided by the Canada First Research Excellence Fund (CFREF), the Natural Sciences and Engineering Research Council of Canada (NSERC), the Canada Excellence Research Chairs (CERC) Program, the Canadian Institute for Advanced Research (CIFAR), the province of Ontario and Industry Canada.} L.Z and C.R acknowledge support from the NSF under Grant No. CHE-1410504.


\end{document}